\documentclass[prd,nofootinbib,preprint,superscriptaddress,secnumarabic]{revtex4}
\pdfoutput=1

\ifx\pdfoutput\undefined
\usepackage[dvips,bookmarks=false]{hyperref}	
\else
\usepackage{hyperref}	
\fi
\hypersetup{colorlinks,bookmarksopen,bookmarksnumbered,citecolor=blue,
linkcolor=blue,pdfstartview=FitH,urlcolor=blue}

\usepackage{graphicx}
\usepackage{amsmath}
\usepackage{amssymb}

\newcommand{\be}{\begin{equation}}
\newcommand{\ee}{\end{equation}}
\newcommand{\beq}{\begin{equation}}
\newcommand{\eeq}{\end{equation}}

\newcommand{\vect}[1]{\boldsymbol{\rm #1}}

\newcommand{\capdef}{}
\newcommand{\mycaption}[2][\capdef]{\renewcommand{\capdef}{#2}%
       \caption[#1]{{\footnotesize #2}}}
\makeatletter
\renewcommand{\fnum@table}{\textbf{\tablename~\thetable}}
\renewcommand{\fnum@figure}{\textbf{\figurename~\thefigure}}
\makeatother
\newcommand{\iso}[2]{{\ensuremath{{}^{#2}}\ensuremath{\rm #1}}}
\newcommand{\bs}[1]{{\ensuremath{\boldsymbol{#1}}}}

\begin{document}
\pagestyle{plain}

\vspace*{1cm}

\title{Light Dark Matter in the light of CRESST-II}
\vspace*{1cm}

\def\Fermilab{Theoretical Physics Department, Fermilab, PO Box 500, Batavia, IL 60510, USA}
\author{\textbf{Joachim Kopp}\vspace*{0mm}}
\email{jkopp AT fnal.gov}
\affiliation{\Fermilab}

\def\mpi{Max-Planck-Institut f{\"u}r Kernphysik, PO Box 103980, 69029 Heidelberg, Germany}
\author{\textbf{Thomas Schwetz}\vspace*{0mm}}
\email{schwetz AT mpi-hd.mpg.de}
\affiliation{\mpi}

\def\Cincy{Department of Physics, University of Cincinnati, Cincinnati, Ohio 45221,USA}
\author{\textbf{Jure Zupan\footnote{On leave of absence from University of Ljubljana, Depart. of Mathematics and Physics, Jadranska 19, \\[-1mm]
 1000 Ljubljana, Slovenia and Josef Stefan Institute, Jamova 39, 1000 Ljubljana, Slovenia.}}\vspace*{3mm}}
\email{jure.zupan AT cern.ch}
\affiliation{\Cincy}

\begin{abstract}
\vspace*{5mm} 
Recently the CRESST collaboration has published the long anticipated results of their direct Dark Matter (DM) detection experiment with a $\rm CaWO_4$ target. The number of observed events exceeds known backgrounds at more than 4$\sigma$ significance, and this excess could potentially be due to DM scattering. We confront this interpretation with null results from other direct detection experiments for a number of theoretical models, and find that consistency is achieved in non-minimal models such as inelastic DM and isospin-violating DM. In both cases mild tension with constraints remain. The CRESST data can, however, not be reconciled with the null results and with the positive signals from DAMA and CoGeNT simultaneously in any of the models we study.
\end{abstract}
\maketitle

\section{Introduction}

At the recent TAUP~2011 conference the CRESST collaboration has reported a significant ($> 4\sigma$) excess of nuclear recoil events in their detector, which cannot be explained by any known source of background~\cite{Angloher:2011uu}. If interpreted in terms of dark matter--nucleus scattering, these events would point to dark matter (DM) masses roughly between 10~GeV and 60~GeV, and DM--nucleon scattering cross section of order $10^{-43}$--$10^{-40}$~cm$^2$. The CRESST result thus provides a third inconclusive hint for the existence of relatively light dark matter, in addition to the long-standing observation of an annual modulation signal by DAMA~\cite{Bernabei:2010mq}, and the more recently reported event excess and annual modulation signal from CoGeNT~\cite{Aalseth:2011wp}.

It is, however, well known~\cite{Schwetz:2011xm,Farina:2011pw,Fox:2011px,McCabe:2011sr,Angloher:2011uu} that the DAMA, CoGeNT and CRESST results, when interpreted in terms of the simplest dark matter models, are inconsistent among each other, and are moreover in severe conflict with null results from other experiments, such as CDMS~\cite{Ahmed:2009zw,Ahmed:2010wy}, XENON-100~\cite{Aprile:2011hi} and KIMS~\cite{Lee:2007qn,KIMS-TAUP} (for recent phenomenological analyses see Refs.~\cite{Cline:2011uu,Natarajan:2011gz,Fornengo:2011sz,An:2011ck,Rajaraman:2011wf,Cline:2011zr,Frandsen:2011cg,Frandsen:2011ts,Hooper:2011hd, Belli:2011kw,Foot:2011pi,Kopp:2009qt,Kopp:2010su,Kopp:2009et,Arina:2011si,Schwetz:2010gv,Fornengo:2010mk,Buckley:2010ve,Gunion:2010dy,Chang:2010en,Chang:2010pr,Savage:2010tg,Alves:2010pt,Graham:2010ca,Chang:2010yk,Andreas:2010dz,Fitzpatrick:2010em}). The updated KIMS result~\cite{KIMS-TAUP} is of particular interest because KIMS' target material CsI(Tl) contains iodine, which is also one of the target elements used in the DAMA's NaI(Tl) crystals. The null result from KIMS rules out any explanation of the DAMA annual sinusoidal modulation signal in terms of iodine recoils. Thus, if DAMA is actually seeing dark matter, the scattering would have to be dominantly off sodium. 

In this paper, we focus on the CRESST result and discuss possible interpretations of the observed events in terms of dark matter scattering. We confirm and quantify that within the conventional scenario of elastic spin-independent scattering the parameter region required by CRESST is disfavored by constraints from other experiments. Therefore we focus on modified particle physics scenarios, and we show that inelastic scattering or isospin-violating interactions may alleviate the tension between CRESST and the null results, though a joint explanation with DAMA and/or CoGeNT seems not to be possible. Let us note that the naturally most abundant isotopes in CRESST's $\rm CaWO_4 $ target have zero spin, and therefore CRESST cannot efficiently probe spin-dependent dark matter interactions.

The outline of the paper is as follows. In sec.~\ref{sec:models} we review the relevant equations for DM direct detection (see also \cite{Fairbairn:2008gz,Kopp:2009qt,Schwetz:2011xm} for more details) and discuss the particle physics scenarios we are going to use to describe the data.
In sec.~\ref{sec:data} we list the data we are going to analyse and provide some details on how we do the fitting of the various experimental data sets. In sec.~\ref{sec:eSI} we consider conventional elastic spin-independent scattering and confirm the severe tension between CRESST and other experiments (sec.~\ref{sec:eSI-cresst}). We also consider, in sec.~\ref{sec:cogent}, the implications of a (possibly large) contamination of the CoGeNT event excess due to inefficient rejection of background events occuring close to the surface of the detector~\cite{Collar-TAUP}. In sec.~\ref{sec:dama} we discuss the possible impact of new preliminary measurements of the quenching factor for sodium recoils in DAMA's target material NaI(Tl)~\cite{Collar-TAUP}. Next, we turn to more exotic particle physics models of DM, considering inelastic scattering (sec.~\ref{sec:iDM}) and isospin-violating interactions (sec.~\ref{sec:IVDM}). In both cases we show that a description of CRESST data consistent with all null results is possible. We conclude in sec.~\ref{sec:conclusions}.

\section{Direct detection and dark matter models}
\label{sec:models}

The differential rate for a DM particle $\chi$ to scatter elastically in a detector composed of nuclei with charge $Z$ and mass number $A$ is given by
\begin{align}
  \frac{dR}{dE_d} = \frac{\rho_\chi}{m_\chi} \frac{1}{m_A} 
    \int_{|\bs{v}| > v_{\rm min}} d^3v \frac{d\sigma_A}{d E_d} \, v f(\bf{v}) \,.
  \label{Eq:rate}
\end{align}
The units in which we will quote $dR/dE$ are events/keV/kg/day. Here $\rho_\chi$ is the local DM density for which we adopt the standard value of $\rho_\chi=0.3 {\rm \,GeV\,cm}^{-3}$. All our cross sections are understood relative to this reference value. More realistic DM densities (see e.g.~\cite{Catena:2009mf}) lead to a trivial re-scaling of our results. $f(\bf{v})$ is the DM velocity distribution in the rest frame of the detector. It is obtained from the velocity distribution in the galactic rest frame, for which we assume a
Maxwellian distribution with a smooth cut-off, $f_\mathrm{gal}(\vect{v}) \propto [\exp(-\vect{v}^2/\bar{v}^2) - \exp(v_{\rm esc}^2/\bar{v}^2)]\Theta(v_{\rm esc}-v)$, where $\bar v = 220\,\rm km\, s^{-1}$ and $v_{\rm esc} = 550\,\rm km\, s^{-1}$  (for more details see \cite{Kopp:2009qt}). The lower limit of the integration in \eqref{Eq:rate} is set by the minimal velocity $v_{\rm min}$ that the incoming DM particle has
to have in order to be able to deposit an energy $E_d$ in the detector. For the case of inelastic scattering, $\chi A\to \chi' A$, it is given by
\begin{align}
  \label{vmin}
  v_{\rm min}=\frac{1}{\sqrt{2m_A E_d}}\left(\frac{m_A
    E_d}{\mu_{\chi A}}+\delta\right),
  \end{align}
where $\delta \equiv m_{\chi'} - m_\chi \ll m_\chi$. The same equation also applies to elastic scattering, with $\delta=0$. 

In the scattering rate \eqref{Eq:rate} the information about the particle physics DM model is encoded in the cross section $d\sigma_A/dE_d$. Since DM is nonrelativistic, the cross section can in general be divided into contributions from couplings to the mass of the nucleus---giving the spin independent (SI) cross section---and couplings to the spin of the nucleus---giving the spin dependent (SD) contribution to the cross section
\begin{align}
  \frac{d\sigma}{dE_d} = \frac{m_N}{2\mu_{\chi N}^2 v^2}\left(\sigma^{\rm SI} F^2(E_d)
                          +\sigma^{\rm SD} S(E_d)\right).
  \label{eq:dsigmadE}
\end{align}
The SI cross section for DM-nucleus scattering can be written in terms of the SI cross section for DM--\emph{nucleon} scattering: 
\begin{align}
  \sigma^{\rm SI}=\frac{[Zf_p+(A-Z)f_n]^2}{f_p^2}
    \frac{\mu_{\chi N}^2}{\mu_{\chi p}^2}\sigma_p^{\rm SI} \,.
  \label{eq:sigmaSI}
\end{align}
Here, $f_{p,n}$ is the normalized SI DM coupling to protons and neutrons, respectively, $\mu_{\chi p}$ is the reduced mass of the DM--nucleon system, and $\sigma_p^{SI}$ is the SI cross section for scattering of DM on a proton. For the SI form factor $F(E_d)$ we use~\cite{Jungman:1995df}
$F(E_d) = 3 e^{-\kappa^2 s^2/2} [\sin(\kappa r)-\kappa r\cos(\kappa
r)] / (\kappa r)^3$, with $s = 1$~fm, $r = \sqrt{R^2 - 5 s^2}$, $R =
1.2 A^{1/3}$~fm, $\kappa = \sqrt{2 m_N E_d}$ (and
$q^2\simeq-\kappa^2$). In this paper, we will not consider spin-dependent scattering (i.e.\ we set $\sigma^{\rm SD} = 0$ in eq.~\eqref{eq:dsigmadE}) since the isotopes which are present in the CRESST detector either have no spin, or have a very small natural abundance, so that CRESST is not very sensitive to spin-dependent DM interactions.

We will analyze the following particle physics models for DM--nucleus interactions:
\begin{itemize}
    \item{\bf Elastic spin independent (eSI)  DM scattering}. This type of DM--nucleus interaction can for instance arise from exchange of a scalar particle (such as a Higgs boson or a squark), or from vector interactions (such as the vectorial part of $Z$ exchange). It is thus very common in many models, examples include supersymmetric neutralino DM~\cite{Jungman:1995df}, the lightest Kaluza-Klein particle in universal extra dimensions~\cite{Cheng:2002ej,Servant:2002hb}, and little Higgs models \cite{BirkedalHansen:2003mpa}. Because DM is nonrelativistic and the momentum transfer in DM--nucleus scattering is small compared to typical nuclear energy scales, eSI scattering occurs coherently off all nucleons in a target nucleus, so that the cross section is enhanced by a factor $A^2$ (the effect of nonzero momentum exchange is accounted for by the nuclear form factors). Conventionally, eSI scattering is assumed to be isospin-conserving, $f_n=f_p=1$. For instance, in models where DM scatters through Higgs exchange, this is because the process predominantly probes the sea quark content of the nucleon, which is the same for protons and neutrons. We will relax the assumption of isospin-independent scattering below. 
    
    \item {\bf Inelastic DM (iDM)}. In iDM models, the DM sector is composed of at least two particles with mass splitting $\delta>0$, so that in a scattering process on a nucleus, the lighter state is converted into the heavier one. In this case, it is possible that DM scattering can give a signal above the typical $\mathcal{O}(\text{10~keV})$ detection thresholds only when the target nuclei are relatively heavy and the DM velocity is high, close to the galactic escape velocity $v_{\rm esc}$. This is required for the momentum exchange to be large enough to overcome the splitting~\cite{TuckerSmith:2001hy}. iDM therefore avoids constraints from searches performed on lighter nuclei, and the annual modulation of the rate due to the change in relative velocity between the Earth and the galactic DM halo is enhanced. The possibility of exothermic scattering, $\delta<0$, has also been discussed in the literature~\cite{Graham:2010ca}. 
      
      \item {\bf Isospin-violating DM (IVDM) scattering}. In general DM can couple differently to protons and neutrons, so that $f_p\ne f_n$ \cite{Giuliani:2005my}.  Note that even DM interactions through Higgs exchange in supersymmetric models are isospin-violating because of the different up and down Yukawa couplings, and only appear isospin-conserving in direct detection experiments because couplings to sea quarks are dominant there. In the presence of new interactions, however, it is quite possible that isospin-violation in DM--nucleus interaction is much stronger. Note that isospin-violating DM does not imply large flavor-changing neutral currents. For instance, there could be different couplings of dark matter to left-handed and right-handed quark vector currents, see e.g.\ the interactions discussed in~\cite{Kamenik:2011nb}. Isospin-violating DM can also be realized in technicolor models, see~\cite{DelNobile:2011je}. If DM couplings to protons and neutrons have opposite sign, $f_p f_n<0$, there can be a cancellation between the two contributions, depending on the number of protons and neutrinos in a given target isotope. For example, for $f_n/f_p\simeq -0.7$ the rate of scattering on Xe is reduced by an order of magnitude \cite{Giuliani:2005my, Chang:2010yk, Feng:2011vu}. We will investigate whether it is possible to use this effect to re-concile the CRESST result with the constraints from other experiments (see also \cite{Frandsen:2011ts,Schwetz:2011xm,Farina:2011pw} for recent analyses in the context of CoGeNT).
\end{itemize}

If the DM interaction is mediated by a particle lighter than the typical momentum transfers in direct detection experiments, of order 100~MeV, the expected energy spectrum will be modified. Recent studies \cite{Schwetz:2011xm,Farina:2011pw,Fornengo:2011sz} have shown that this does not improve the consistency of the DM hints with various constraints and therefore we will not discuss light mediators here.

\section{Fitting direct detection data}
\label{sec:data}

In this section, we discuss the inputs to our fits for the various direct
detection experiments we consider.

\subsection{CRESST}
\label{sec:cresst}
CRESST-II, which uses $\rm CaWO_4$ crystals as target material, obtains after 739 kg days of exposure 67 events that fulfill all acceptance criteria~\cite{Angloher:2011uu}. In order to reject backgrounds, the CRESST detectors record for each event both the energy deposited in phonons and the amount of scintillation light emitted. The scintillation light yield can be used to distinguish the signal from most sources of background, and also to distinguish scattering processes on the different target nuclei, W, Ca and O. The known backgrounds in CRESST are electrons and gamma rays from radioactive decays, $\alpha$ particles, neutrons, and \iso{Pb}{210} decays due to surface contamination. The total deposited energy in each event is measured in the phonon channel, where the lower energy threshold is chosen separately for each of the eight detector modules (see Table~1 in~\cite{Angloher:2011uu}), according to the requirement that the expected number of electro-magnetic background events in the signal window is less than one. The CRESST collaboration has found that only about 65--75\% of the observed events can be accounted for by the known backgrounds, corresponding to an excess with a statistical significance of more than $4\sigma$.

While the CRESST collaboration performs a likelihood fit of DM parameters to their data, using information on the energy deposit and the light yield for each event, we perform a simplified fit utilizing only publicly available information, which does not include the light yield for each event. In particular, we fit the total event rate in each detector module, as well as the overall energy spectrum. We take into account the individual energy threshold of each detector module. Following the CRESST collaboration, we assume backgrounds due to Pb recoils, $\alpha$ particles and neutrons to be constant in energy (see fig.~11 in \cite{Angloher:2011uu}), whereas the $e/\gamma$ background, which is one event per detector module, is assumed to appear in the lowest energy bin of each module. We compare our predicted energy spectrum to the data shown in fig.~11 of~\cite{Angloher:2011uu} using a $\chi^2$ analysis, in which the normalization of the flat background is allowed to float independently for each detector module.

\subsection{CRESST commissioning run}
In an earlier commissioning run of the CRESST
experiment~\cite{Angloher:2008jj} using only two detector modules, the
collaboration has collected a combined exposure of 47.9~kg~days and observed
three candidate events. The energy threshold of 10~keV in that analysis was
lower than in the full run, whereas the upper end of the signal windows at
40~keV was the same. Using updated acceptance factors and considering DM
scattering on all three target elements (the CRESST collaboration only
considered tungsten recoils in that analysis), tight constraints are
obtained, excluding part of the preferred region from the longer CRESST-II
run~\cite{Brown:2011dp}. We include this constraint in our analysis using
the energy and isotope dependent efficiences given in \cite{Brown:2011dp},
and we apply the maximum-gap method~\cite{Yellin:2002xd} to compute an
exclusion limit.

\subsection{CoGeNT}
The CoGeNT collaboration has first reported an excess of low-energy events---possibly due to dark matter---in 2010~\cite{Aalseth:2010vx}. Subsequently, this excess has been confirmed, and it has been found to show a seasonal variation at $2.8\sigma$ significance~\cite{Aalseth:2011wp}. Very recently, new investigations of CoGeNT's backgrounds have indicated the possibility that a yet unknown fraction of the observed excess events is due to activity close to the surface of the detector, where external radioactivity is prevalent. Our analysis of the CoGeNT data from~\cite{Aalseth:2010vx,Collar:2011privcomm} is similar to the one presented in~\cite{Fox:2011px}. We carry out an unbinned extended maximum likelihood fit, thus making the best possible use of the recorded energy of each individual event. In addition to the peaked cosmogenic backgrounds and the flat background considered in ref.~\cite{Fox:2011px}, we also include an exponential component of the form $a \exp(-E / E_0)$ to account for the possible contamination of the event sample by surface events. Since the values of the parameters $a$ and $E_0$ are not known precisely yet, we will consider different possibilities. We do not include timing information in our fit, since it is conceivable that not only the dark matter signal, but also the surface backgrounds show seasonal variation---for instance due to fluctuating levels of environmental radioactivity---and it would be very difficult to disentangle these different sources of annual modulation, especially with the still relatively low statistics in CoGeNT.

\subsection{XENON-100}
In a 4843~kg~day exposure of the XENON-100 detector, three candidate events have been observed, with a background expectation of $1.8\pm 0.6$ events~\cite{Aprile:2011hi}, so there is no evidence for DM scattering. These data lead not only to the world's strongest limit on conventional spin-independent elastic WIMP--nucleon scattering, constraining the corresponding cross section to below $10^{-44} \, {\rm cm}^2$ for $m_\chi \sim 100$~GeV, but have also important implications for the low mass region around 10~GeV (see also the discussion in~\cite{Collar:2010gg, XENON:2010er, Collar:2010gd, Sorensen:2010hq, Manalaysay:2010mb, Collar:2010ht, Bezrukov:2010qa, Sorensen:2011bd, Collar:2011wq}).  We derive a conservative limit by using the maximum gap method from~\cite{Yellin:2002xd}. 

An important input to the limits from xenon experiments is the scintillation light yield, parametrized conventionally by the function $L_{\rm eff}(E_{\rm nr})$, see e.g.~\cite{Savage:2010tg}. We use for $L_{\rm eff}(E_{\rm nr})$ the black solid line from fig.~1 of~\cite{Aprile:2011hi}. It has been obtained from a fit to various data, but is dominated at low energies by the recent measurement from ref.~\cite{Plante:2011hw}. We include an overall systematic uncertainty on $L_{\rm eff}$, which we assume to be $\pm 0.01$ above $E_{\rm nr} = 3$~keV, and increasing linearly by $0.05/\text{keV}$ below. The assumptions on $L_{\rm eff}$ below $E_{\rm nr} \approx 3$~keV (where no data are availble) have no impact on the XENON-100 exclusion curves. Even assuming $L_{\rm eff} = 0$ below 3~keV, the limits remain essentially unchanged~\cite{Schwetz:2011xm}. Hence, the result does not rely on any extrapolation into a region with no data. 

\subsection{KIMS}
The KIMS experiment uses a CsI target, and is thus directly probing whether the DAMA signal can be due to DM scattering on iodine. The KIMS collaboration have recently presented results based on 24524.3~kg~days of exposure, corresponding to roughly 1~year of data taking~\cite{KIMS-TAUP}. Since they do not see any events at the relevant recoil energies below 8~keVee\footnote{The units keVee are used for ``electron equivalent energy'', i.e.\ the amount of energy deposited in electronic excitations. Usually, this is only a fraction (given by the quenching factor) of the initial nuclear recoil energy, for which we use the unit keVnr.}, they are able to exclude the possibility that the DAMA annual modulation can be due to WIMPs recoiling on iodine. Note that this statement is very robust with respect to astrophysical uncertainties and our ignorance of the underlying particle physics model of DM.  KIMS results also place strong constraints on inelastic DM because both Cs and I are heavy, so that iDM scattering is very efficient in KIMS.

\subsection{CDMS-II}
CDMS-II bounds are derived from the 612~kg~days of data taken with Ge detectors in four periods between July 2007 and September 2008~\cite{Ahmed:2009zw}, and from the previous CDMS search which had an exposure of 397.8~kg~days, obtained between October 2006 and July 2007~\cite{Ahmed:2008eu}. In~\cite{Ahmed:2009zw} two events with  recoil energies of 12.3~keVnr and 15.5~keVnr were observed in the 10-100~keVnr signal window. The expected backgrounds from misidentified surface events, cosmogenic backgrounds and neutrons were $0.8\pm0.1\pm0.2, 0.04^{+0.04}_{-0.03}$ and 0.03--0.06, respectively. In our fits we use again Yellin's maximum gap method~\cite{Yellin:2002xd} to set limits. We take into account a linear efficiency drop from 32\% at 20~keVnr to 25\% at both 10~keVnr and 100~keVnr, and we use a constant energy resolution of 0.2~keV. 

\subsection{CDMS-Si and CDMS low threshold analysis}
A stringent constraint on low mass DM is obtained from a modified reanalysis of the CDMS-II data collected in eight germanium detectors between October 2006 and September 2008~\cite{Ahmed:2010wy}. In this analysis, the CDMS collaboration lowered the analysis energy threshold to 2~keVnr at the expense of allowing for a larger background contamination, and obtained and enhanced sensitivity to DM--nucleus scattering for WIMP masses below $\sim 10$ GeV. In our analysis of the low-threshold data, we use the $\Delta\chi^2$ method, but include only those bins in which the predicted number of events is larger than the observed one. The CDMS collaboration has also collected 12~kg~days of data on silicon targets \cite{Akerib:2005kh}. No signal events above expected background were observed. Since silicon is lighter than germanium this places competitive bounds on light DM, especially in the case of isospin-violating DM models. 

\subsection{DAMA/LIBRA}
The combined exposure of the DAMA/LIBRA experiment and the previous DAMA/NaI  experiment (DAMA for short) is 1.17 ton yr and spans 13 annual cycles \cite{Bernabei:2010mq}. The data show an annual modulation signal at 8.9$\sigma$ significance, with a peak around the end of May. This is consistent with the expectation from WIMP scattering, which predicts a peak around June 2nd for standard assumptions on the DM halo. In the fit we use the signal region from 2~keVee to 8~keVee. Above this range the data are consistent with no modulation. Furthermore, no signal is predicted above this range from the DM models considered, so higher energy recoils give no additional constraint on the fit and can be ignored. Our analysis of the DAMA data is analogous to the one presented in~\cite{Fairbairn:2008gz,Kopp:2009et,Kopp:2009qt}. Unless otherwise noted, we use quenching factors $q_\mathrm{Na} = 0.3 \pm 0.03$ and $q_\mathrm{I} = 0.09 \pm 0.009$~\cite{Bernabei:1996vj} and ignore channeling effects, which are expected to be small~\cite{Bozorgnia:2010xy}. Note that we include independent 10\% systematic uncertainties on $q_\mathrm{Na}$ and $q_\mathrm{I}$.

\section{Elastic spin independent scattering}
\label{sec:eSI}

\subsection{Fitting CRESST results}
\label{sec:eSI-cresst}

\begin{figure}
  \includegraphics[width=0.55\textwidth]{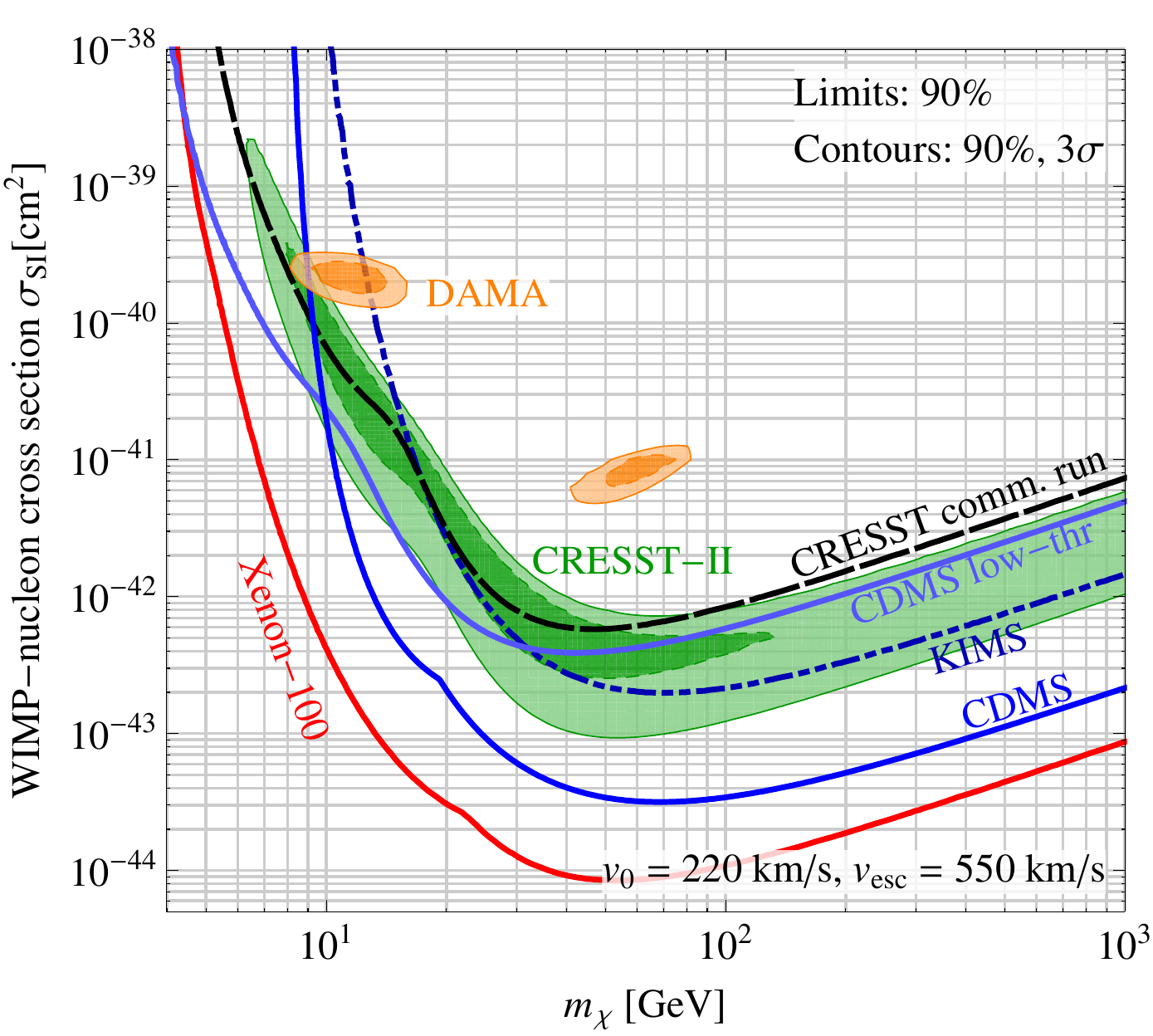}
  \mycaption{Constraints on elastic, spin-independent, isospin-conserving
    DM--nucleon scattering.  We show the parameter regions preferred by the
    CRESST-II and DAMA signals (for CoGeNT see fig. \ref{fig:CoGeNTfloat}),
    together with constraints from KIMS, CDMS (high threshold and low threshold
    analyses), XENON-100 and the CRESST commissioning run.}
  \label{fig:eSI}
\end{figure}

We first assume that the signal in CRESST is due to DM scattering elastically
with spin independent couplings (eSI). The resulting allowed regions in DM mass
$m_\chi$ and DM--proton scattering cross section $\sigma_p$ assuming equal
couplings to protons and neutrons ($f_p=f_n$) are shown in fig.~\ref{fig:eSI}.
We obtain our best fit point at $m_\chi = 12.5$~GeV and $\sigma_p = 2.7\times
10^{-41} \,\rm cm^2$, while the degenerate solution with somewhat higher mass
and smaller cross section, $m_\chi=29$ GeV and $\sigma_p=10^{-42} \,\rm cm^2$,
gives a description of data that is only slightly worse by $\Delta \chi^2=0.3$.
The agreement between our regions and the ones obtained by the CRESST
collaboration validates our simplified procedure. As in \cite{Angloher:2011uu}
we find two local minima of $\chi^2$ at masses around 12~GeV and 25~GeV. Since
we cannot use the information on the scintillation light distribution our fit
is slightly less constraining than the CRESST collaboration's, and the
parameter regions corresponding to the two local minima are actually merged
into one region at 90\%~confidence level. Note that for small DM masses,
scattering on W does not contribute since the attainable recoil energies on
such a heavy target are below threshold. For higher DM masses, W becomes more
important; for instance, the fraction of tungsten recoils reaches 90\% for
$m_\chi \simeq 50$~GeV~\cite{Angloher:2011uu}. The CRESST region for low DM
masses has some overlap with the region where scattering on Na can explain the
modulation signal in DAMA, under the assumption of a quenching factor $q_{\rm
Na} = 0.3 \pm 0.03$ in DAMA.

It has been noted in \cite{Angloher:2011uu} and stressed in
\cite{Brown:2011dp} that data from the CRESST commissioning run
\cite{Angloher:2008jj} imposes relevant constraints in the parameter region
preferred by currect CRESST data. The limit from the commissioning run is
shown in fig.~\ref{fig:eSI} as a black long-dashed curve, which
excludes part of the region preferred by latest data, but a consistent
region remains even at 90\%~CL. Since both runs use very similar detectors
(up to minor differences in acceptances, backgrounds and energy thresholds),
it is very difficult to alleviate the tension by invoking exotic particle
physics.  Since the commissioning run took place between the end of March
and the end of July 2007 it is conceivable that non-standard DM halos with
strong seasonal variation in the scattering rate could reduce the tension.
Note that the recent CRESST analysis and the commissioning run data are
based on different acceptance cuts, and a direct comparison might be
subject to systematic uncertainties.

In order to quantify agreement or disagreement between data sets, we use the
parameter goodness of fit (PG) test~\cite{Maltoni:2003cu}. This test is
based on the $\chi^2$ function 
\begin{align}
  \chi^2_\text{PG} = \Delta\chi^2_1 + \Delta\chi^2_2 
    \qquad\text{with}\qquad
  \Delta\chi^2_i = \chi^2_i(\text{global bf}) - \chi^2_{i,\rm min} \,,
\end{align}
where the index $i = 1,2$ labels the data sets whose compatibility is to be
tested, and $\Delta\chi^2_i$ is the difference between the $\chi^2$ of the
$i$-th data set at the global best fit point (i.e., at the minimum of
$\chi^2_1 + \chi^2_2$) and the minimum $\chi^2$ from a fit to the $i$-th data
set alone. $\chi^2_\text{PG}$ measures the ``price'' one has to pay for
combining the data sets, compared to fitting them independently.  $\chi^2_\text{PG}$
follows a $\chi^2$ distribution with a number of degrees of freedom
corresponding to the number of parameters to which both data sets are sensitive
(see~\cite{Maltoni:2003cu} for a precise definition). As shown in
Table~\ref{tab:PG}, the PG test finds consistency between the full CRESST-II data
set and the data from the commissioning run at the level of 10\%. The combined
best fit point is obtained at $m_\chi = 12.9$~GeV and $\sigma_p = 2.0\times
10^{-41} \,\rm cm^2$.

\begin{table}
  \begin{ruledtabular}
  \begin{tabular}{lccccccc}
    Data set & $\chi^2_\text{min}$ & $m_\chi$ [GeV] & $\sigma \, [\text{cm}^2]$ & $\delta$ [keV]& $f_n/f_p$
        & $\Delta\chi^2\ [\chi^2_\text{PG}]$ & PG \\
    \hline
    \multicolumn{8}{c}{\bf Elastic spin-independent scattering (eSI)} \\
    \hline
    CRESST & 27.7 & 12.5 & $2.7\times 10^{-41}$ & 0 & 1 & 0.3\\ 
    CRESST$_\text{com}$ & 0 & 0 & 0 & 0 & 1 & 4.2\\ 
    Combination & 32.2 & 13.2 & $2.0\times 10^{-41}$ & 0 & 1 & [4.5] & 10\% \\ 
    \hline
    CRESST & 27.7 & 12.5 & $2.7\times 10^{-41}$ & 0 & 1 & 25.8\\
    Xe100+CDMS$_\text{LT}$+CRESST$_\text{com}$ & 0 & 0 & 0 & 0 & 1 & 0.00\\ 
    Combination & 53.5 & 0.8 & $7.9\times 10^{-45}$ & 0 & 1 & [25.8] & $<10^{-5}$ \\
    \hline

    \multicolumn{8}{c}{\bf Inelastic spin-independent scattering (iDM)} \\
    \hline
    CRESST+CRESST$_\text{com}$ & 32.2 & 13.2 & $2.0\times 10^{-41}$ & 0 & 1 & 7.6\\
    Xe100+CDMS$_\text{LT}$ & 0 & 0 & 0 & 0 & 1 & 1.9\\
    Combination & 41.9 & 34.9 & $1.7\times 10^{-38}$ & 94 & 1 & [9.5] & 2\% \\
    \hline

    \multicolumn{8}{c}{\bf Isospin-violating dark matter (IVDM)} \\
    \hline
    CRESST+CRESST$_\text{com}$ & 31.9 & 14.2 & $3.9\times 10^{-40}$ & 0 &$-0.66$ & 4.1\\
    Xe100+CDMS+CDMS$_\text{Si}$ & 0 & 0 & 0 & 0 & 1 & 9.2\\
    Combination & 45.1 & 14.2 & $2.6\times 10^{-40}$ & 0 & $-0.70$  & [13.3] & 0.4\% \\ 
  \end{tabular}
  \end{ruledtabular}
  \mycaption{Consistency of different experimental data sets for various particle
    physics models using the parameter goodness-of-fit test~\cite{Maltoni:2003cu}.
    We list the minimal $\chi^2$ for the different data sets and for their
    combinations, as well as the dark matter masses and dark matter--proton
    scattering cross sections at these minima. The column labeled
    ``$\Delta\chi^2\ [\chi^2_\text{PG}]$'' lists, for the indvidual data sets,
    the $\Delta\chi^2$ of the combined best fit point with respect to the
    individual best fit points, and $\chi^2_\text{PG} \equiv \Delta\chi^2_1 +
    \Delta\chi^2_2$ for the combined analyses. In the column ``PG'' we give the
    probability of consistency from the parameter goodness of fit test, which
    is obtained by evaluating the cumulative distribution function of the
    $\chi^2$ distribution with the appropriate number of degrees of freedom (2
    for eSI and 3 for iDM and IVDM, see text for details) at $\chi^2_\text{PG}$.}
  \label{tab:PG}
\end{table}

For comparison we show in fig.~\ref{fig:eSI} also constraints imposed on the
eSI DM mass and cross section by various null searches, confirming that an
interpretation of CRESST data in terms of elastically scattering
spin-independent and isospin-conserving dark matter is ruled out by XENON-100
\cite{Aprile:2011hi}, CDMS \cite{Ahmed:2009zw}, and the CDMS low threshold
analysis \cite{Ahmed:2010wy}. As we can see from Table~\ref{tab:PG}, the PG
test gives a probability for consistency between CRESST versus CDMS and XENON
of less than $10^{-5}$.

Below,we discuss modified particle physics models with the aim of bringing CRESST
results into agreement with those bounds. Before we do that, however, let us
briefly address recent developements concerning CoGeNT and DAMA.

\subsection{Increased background in CoGeNT?}
\label{sec:cogent}

\begin{figure}
  \begin{tabular}{cc}
    \includegraphics[width=0.45\textwidth]{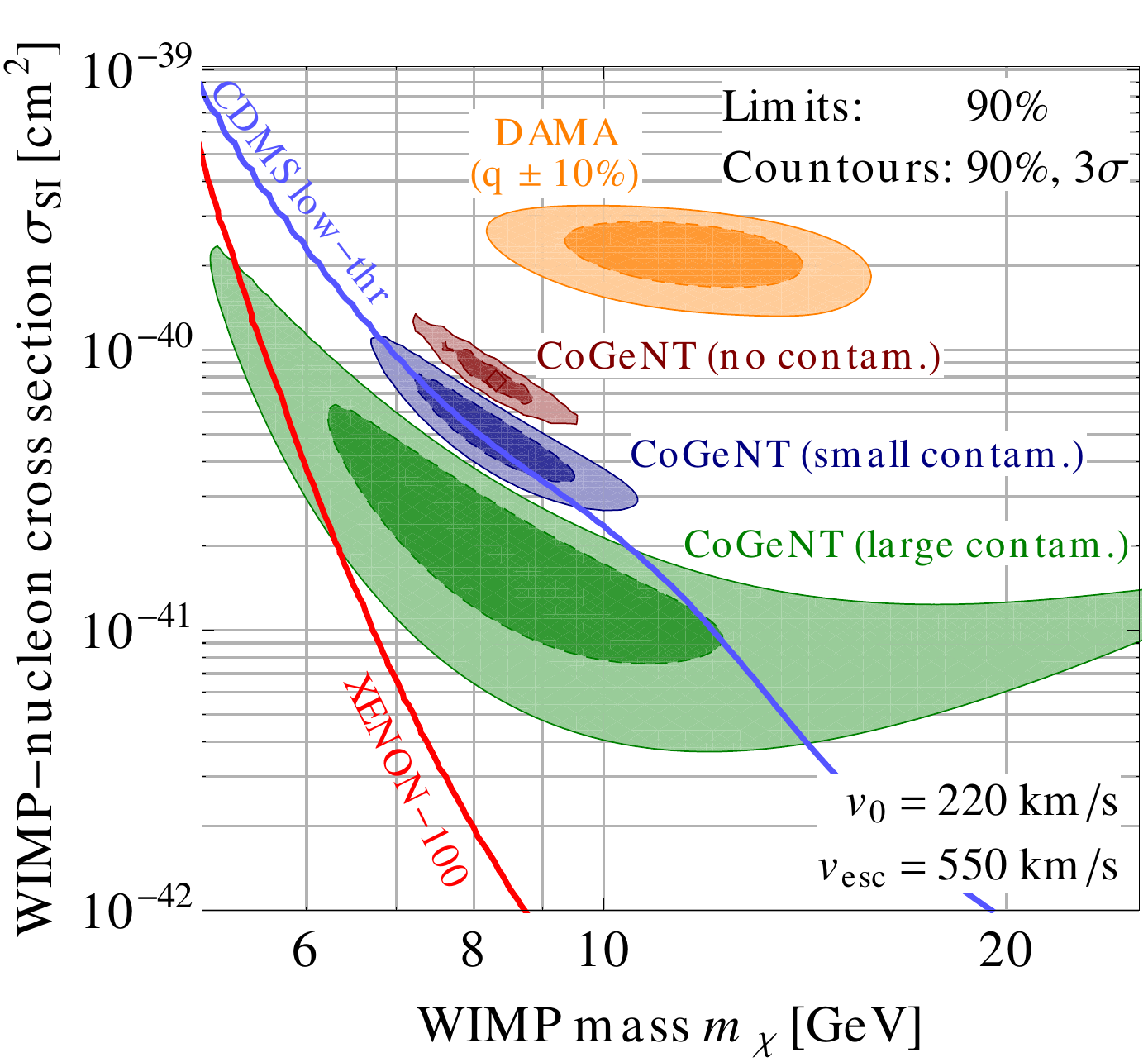} \hspace{0.2cm} &
    \includegraphics[width=0.45\textwidth]{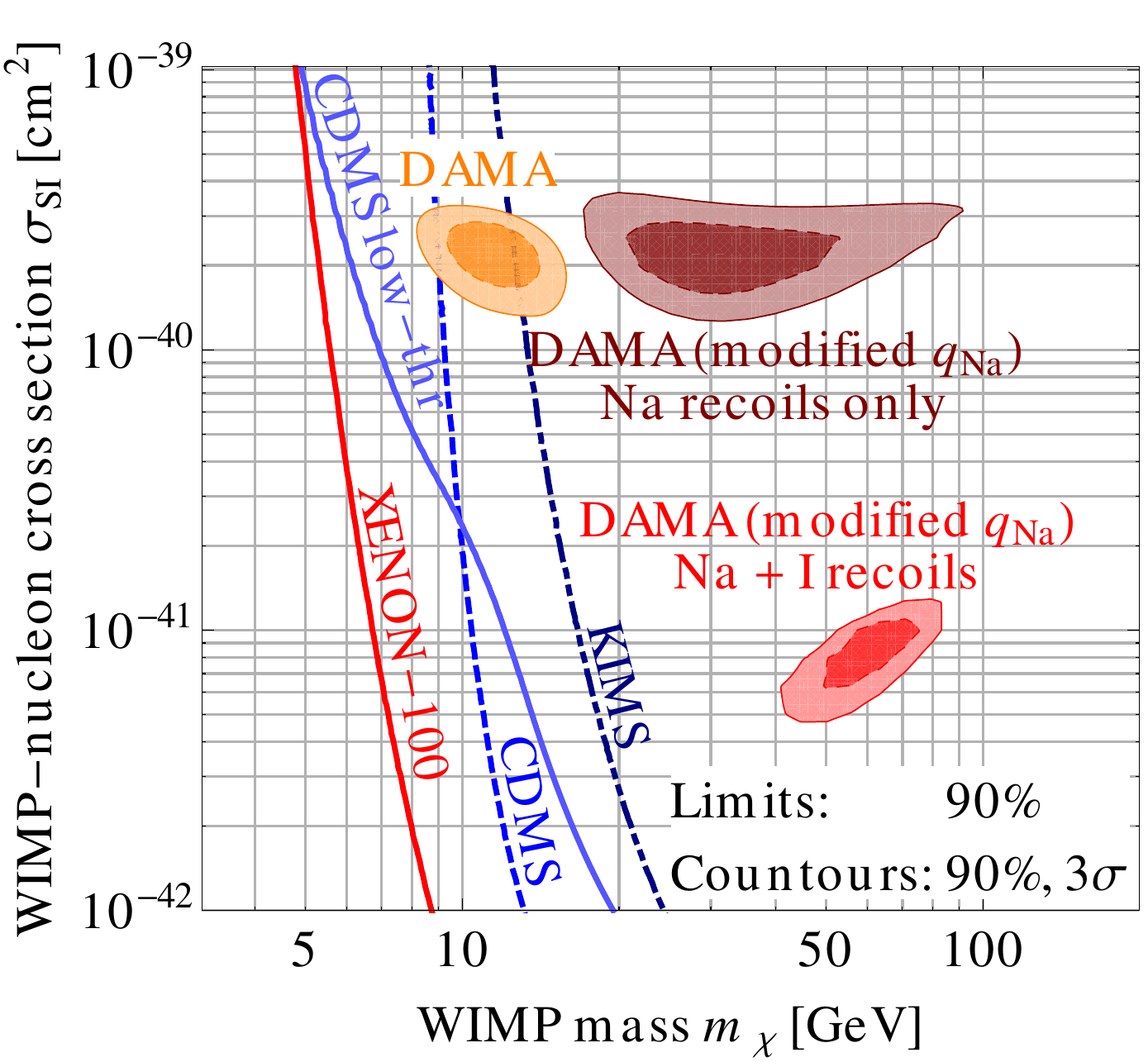} 
  \end{tabular}
  \mycaption{Left: Variation in the CoGeNT-favored parameter region for elastic spin-independent dark matter--nucleon scattering for different assumptions on the surface background in CoGeNT (see text for details). Right: The DAMA preferred regions for the standard assumption on the sodium quenching factor, $q_{\rm Na}=0.3$, (orange region), for the smaller energy dependent $q_{\rm Na}$ obtained in~\cite{Collar-TAUP} (light red region), and for the hypothetical case where scattering on iodine is forbidden (dark red region). In all cases, we have assumed a 10\% overall systematic uncertainty on the DAMA quenching factors.}
  \label{fig:CoGeNTfloat}
\end{figure}

A recent reanalysis of CoGeNT data has revealed the possibility that a significant fraction of the so far unexplained low energy event excess is due to contamination by background activity on the surface of the detector (``surface events'')~\cite{Collar-TAUP}. We investigate in fig.~\ref{fig:CoGeNTfloat} (left) what the implications of this increased background are for the deduced properties of DM (assuming the remaining signal is due to elastic, isospin-conserving, spin-independent dark matter--nucleus scattering). We parameterize the surface background as $a \exp(-E/E_0)$, where $E_0$ is kept fixed at a value of 0.3~keVee, as suggested by the plots in~\cite{Collar-TAUP}. Since the rate of surface events in the signal sample has not been determined precisely yet, we show results for $a = 0$~keV$^{-1}$ day$^{-1}$ (red), $a = 12$~keV$^{-1}$ day$^{-1}$ (blue), and $a = 24$~keV$^{-1}$ day$^{-1}$ (green). At the respective best fit points in $m_\chi$ and $\sigma_p$, these values correspond to signal fractions in the 0.5--1.0~keVee interval of about 73\%, 46\% and 30\%, respectively. Fig.~\ref{fig:CoGeNTfloat} (left) shows that, as expected, larger background contamination shifts the CoGeNT-preferred region in the $m_\chi$--$\sigma$ plane towards lower cross sections and increases its overall size. Similar conclusions have been reached in~\cite{Collar-TAUP}.

\subsection{New sodium quenching factors for DAMA?}
\label{sec:dama}

The quenching factors in the DAMA experiment, which are required to convert the observed energy deposit (in units of keVee) to the true nuclear recoil energy from a dark matter interaction (in keVnr), have been under active discussion for some time already~\cite{Bernabei:2007hw,Bozorgnia:2010xy,Hooper:2010uy}, and a new facet has been added to this discussion recently, when Collar et al.\ carried out an independent measurement of $q_{\rm Na}$, the quenching factor for sodium recoils in NaI(Tl)~\cite{Collar-TAUP}. Their results, which are in tension with results from other groups \cite{Chagani:2008in,Spooner:1994ca,Tovey:1998ex,Gerbier:1998dm,Simon:2002cw}, indicate that $q_{\rm Na}$ may be lower than the standard value $q_{\rm Na} = 0.3$ used by the DAMA collaboration and in many phenomenological analysis. In particular, Collar et al.\ find $q_{\rm Na}\simeq 0.1$ at 30~keVnr nuclear recoil energy and $q_{\rm Na}\simeq 0.2$ at 200~keVnr. This pronounced energy dependence of $q_{\rm Na}$ has also not been seen in previous measurements.

In fig.~\ref{fig:CoGeNTfloat} (right), we explore the implications that Collar et al.'s results, should    they be confirmed, would have on the parameter space of elastic spin-independent DM--nucleon scattering. The orange regions show the parameter values preferred by DAMA at the 90\% and $3\sigma$ confidence levels for the standard assumption $q_{\rm Na} = 0.3 \pm 0.03$, $q_{\rm I} = 0.09 \pm 0.009$. (As before,  we include a 10\% systematic uncertainty on the quenching factors.) In this parameter region, the signal in DAMA is dominated by Na recoils since only a small fraction of $\mathcal{O}(10\text{ GeV})$ dark matter particles can transfer enough energy to an iodine nucleus to induce an event above the experimental energy threshold. The light red contours in fig.~\ref{fig:CoGeNTfloat} (right) show the preferred region one obtains when the sodium quenching factors are taken from~\cite{Collar-TAUP}. Note that at the relevant DM masses $\sim 60$~GeV, the signal in DAMA would be dominated by iodine recoils. For comparison, we also show, in dark red, where the DAMA-allowed region would lie in a hypothetical model where iodine recoils are switched off. (Such a scenario could be realized, for  instance, in the framework of resonant DM~\cite{Bai:2009cd} or isospin-violating DM~\cite{Feng:2011vu}.) We see that in this case, the cross section required to explain DAMA remains roughly the same as with the standard quenching factors, but the preferred dark matter masses are larger, since more energy is required for an event to be above threshold. We conclude that, if the low values for $q_{\rm Na}$ reported in \cite{Collar-TAUP} were confirmed, an interpretation of the DAMA result in terms of eSI scattering would become even more difficult than it already is for $q_{\rm Na} = 0.3$, since the corresponding allowed region is strongly excluded by contraints from XENON-100, CDMS, KIMS, and others. Note in particular, that there is no longer any region where the DAMA modulation is explained in terms of Na recoils. Iodine recoils, on the other hand, are disfavored in a rather model-independent way by KIMS~\cite{KIMS-TAUP}. For the DAMA regions shown in the remaining figures of this work we stick to the value $q_{\rm Na} = 0.3\pm 0.03$.

\section{Inelastic scattering}
\label{sec:iDM}

\begin{figure}
    \includegraphics[height=0.4\textwidth]{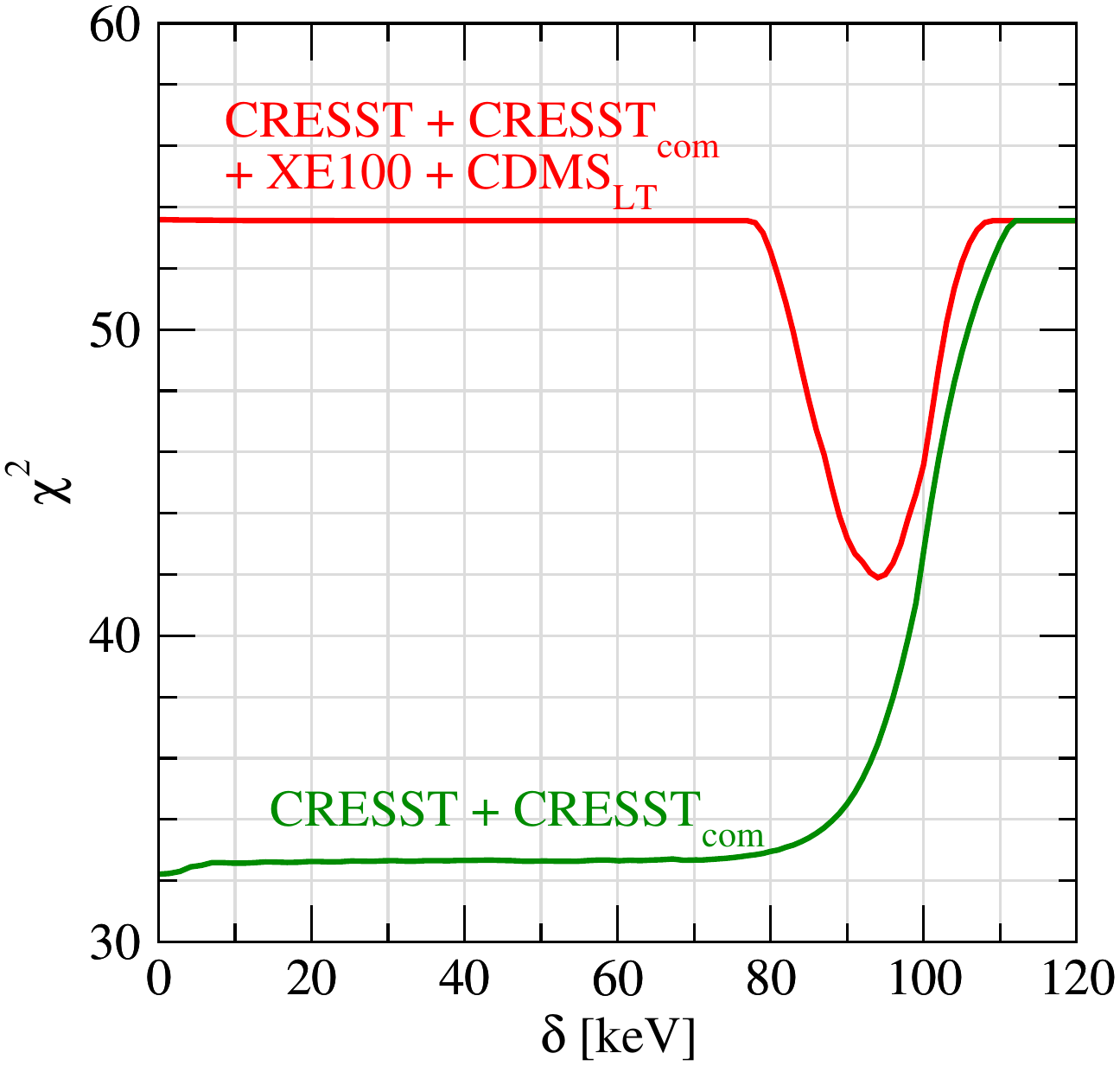} \quad 
    \includegraphics[height=0.42\textwidth]{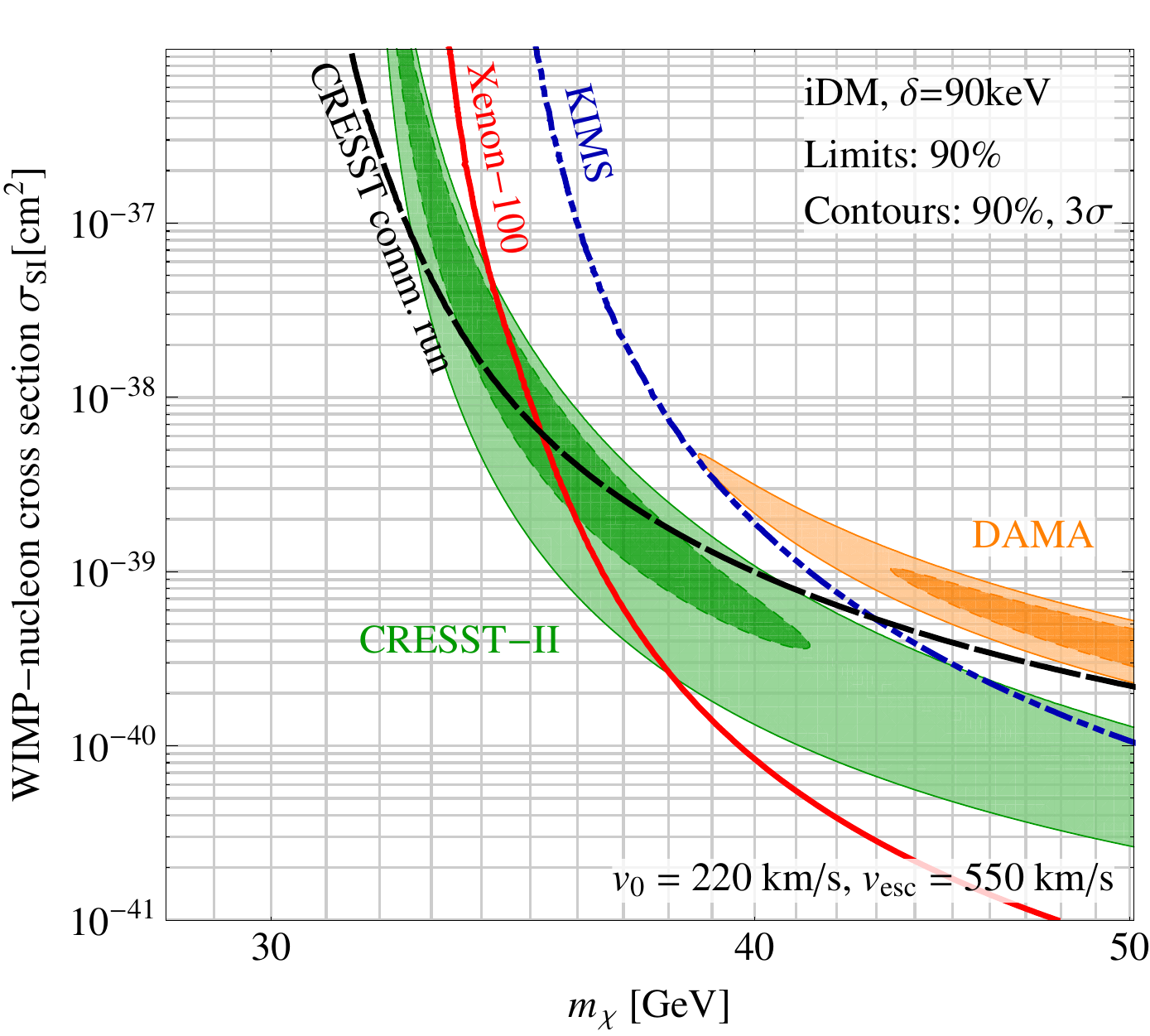}
  \mycaption{Left: the total $\chi^2$ for CRESST (recent data from
    \cite{Angloher:2011uu} combined with the constraint from the commissioning
    run following \cite{Brown:2011dp}) and for CRESST plus data from XENON-100
    \cite{Aprile:2011hi} and CDMS low-threshold \cite{Ahmed:2010wy} as a function
    of the mass splitting $\delta$ for inelastic DM. Right: iDM solution to
    CRESST for $\delta = 90$~GeV, together with constraints from the CRESST
    commissioning run, from KIMS, and from XENON-100. The DAMA preferred region
    for that $\delta$ is also shown.} \label{fig:iDM}
\end{figure}

We next investigate the possibility that the CRESST-II signal is due to
inelastic scattering of DM on nuclei. Fig.~\ref{fig:iDM} (left) shows the
$\chi^2$ of CRESST as well as CRESST combined with XENON and CDMS as a
function of the mass splitting $\delta$. We observe that CRESST alone
prefers elastic scattering, although a reasonable fit is possible up to
$\delta\simeq 95$~keV. However, the combined $\chi^2$ shows a pronounced
minimum around $\delta \simeq 94$~keV. We show the allowed region for CRESST
and the exclusion limits from XENON-100 and KIMS in fig.~\ref{fig:iDM}
(right) for $\delta \simeq 90$~keV. We observe that a CRESST region
consistent with all constraints appears. The consistency according to the
PG test has a probability of about 2\% (see table~\ref{tab:PG}), which is a
huge improvement compared to the elastic case. The still somewhat small
value of this probability indicates that some tension is left between CRESST
and limits from other experiments. 

\begin{figure}
  \includegraphics[width=0.5\textwidth]{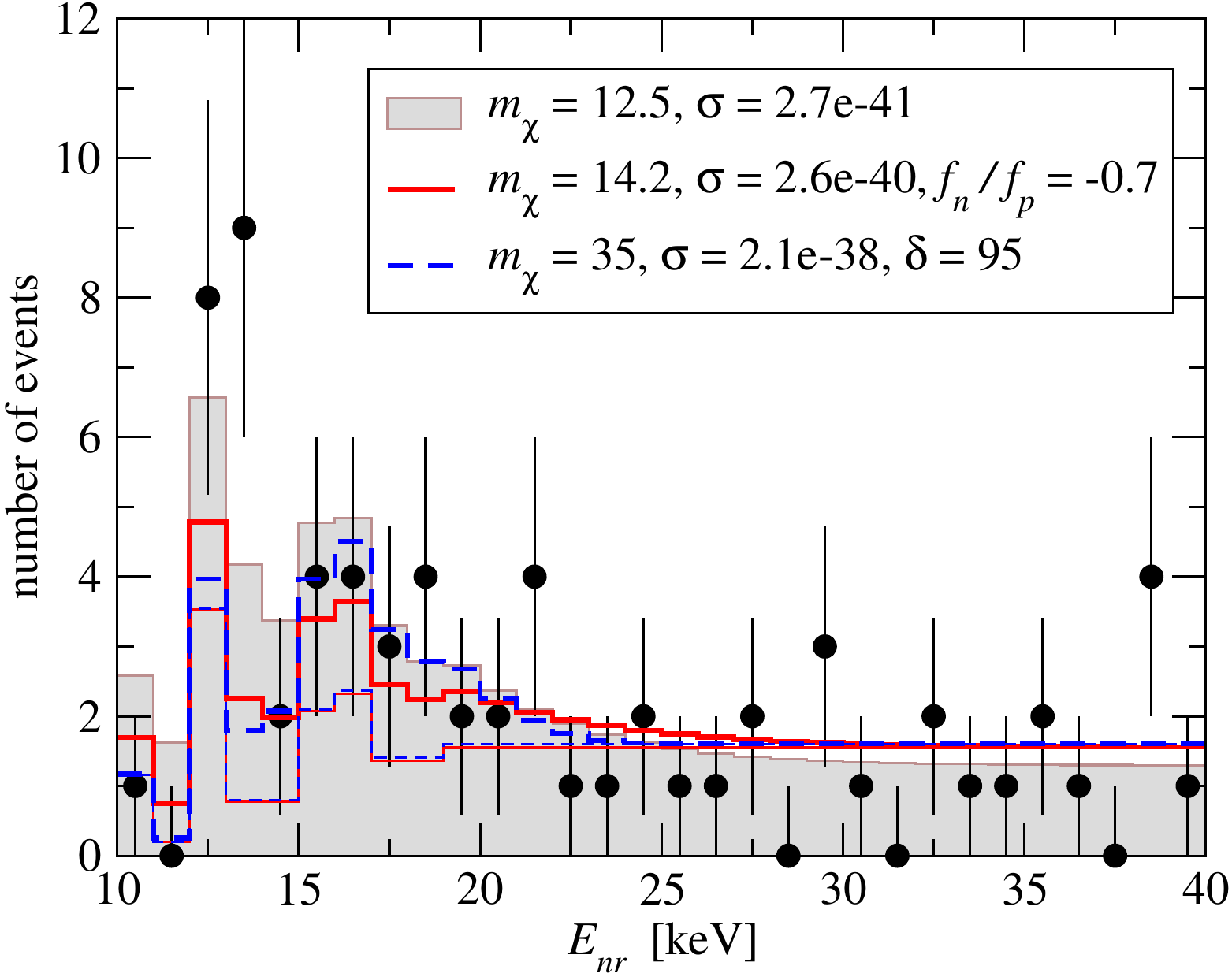}
  \mycaption{The observed CRESST event spectrum compared to predictions
    for a few benchmark models. We show results for elastic SI scattering
    (shaded area), inelastic SI scattering (dashed blue histogram) and
    isospin-violating SI interactions (solid red histogram). The parameter
    values given in the legend are understood to be in GeV for $m_\chi$, cm$^2$
    for the cross section, and keV for $\delta$. Thin curves show the
    background only, while thick curves include both the signal and the
    background.}
  \label{fig:spectrum}
\end{figure}

Inelastic scattering favours heavy target nuclei. Using the very heavy tungsten
nucleus ($A \approx 184$) the rate in CRESST can be enhanced relative to the
rate in all other experiments, which leads to the improved consistency. On the
other hand, for increasing $\delta$ the signal events in an iDM scenario are
shifted to higher energies, and since the CRESST data are more concentrated at
low energies, too high values of $\delta$ are disfavored by the data,
explaining the behaviour of the $\chi^2$ curves in fig.~\ref{fig:iDM}. In
fig.~\ref{fig:spectrum} we show the predicted spectrum at a parameter point
close to the iDM best fit point. We see that, although it provides a reasonable
description of the data, iDM cannot very well account for all of the observed
excess at low energy. The total number of predicted signal events at this
parameter point is 12.2. In comparison we show also the predicted spectrum at
the best fit point for elastic scattering (shaded histogram), where the number
of predicted signal events is 24.8. 

In case of iDM with $\delta \simeq 90$~keV, scattering in CRESST occurs
exclusively on W since scattering on the light O and Ca nuclei is kinematically
forbidden. At the best fit point for elastic spin-independent scattering, with
$m_\chi = 12.5$~GeV, on the other hand, we find a negligible contribution from
W recoils and O (Ca) recoils accounting for 47\% (53\%) of the rate.  While W
recoil events from inelastic DM scattering do provide a reasonable fit to the
$E_{nr}$ spectrum shown in fig.~\ref{fig:spectrum} the question remains whether
also the scintillation light yield distribution is consistent with scattering
exclusively on W. Unfortunately there is not enough information publicly
available to check this explicitly. Let us note however that the favored
parameter regions found by the CRESST collaboration in their fit of elastically
scattering DM~\cite{Angloher:2011uu} contain solutions that correspond to
W-fractions as large as 90\%.  This suggests that because of the large overlap
between the acceptance bands for W, O and Ca, scattering dominantly on W is
consistent with the data, The final answer to this question can be obtained,
however, only by a full fit to the 2-dimensional $E_{nr}$--light yield
distribution.

\section{Isospin-violating scattering}
\label{sec:IVDM}

\begin{figure}
    \includegraphics[height=0.45\textwidth]{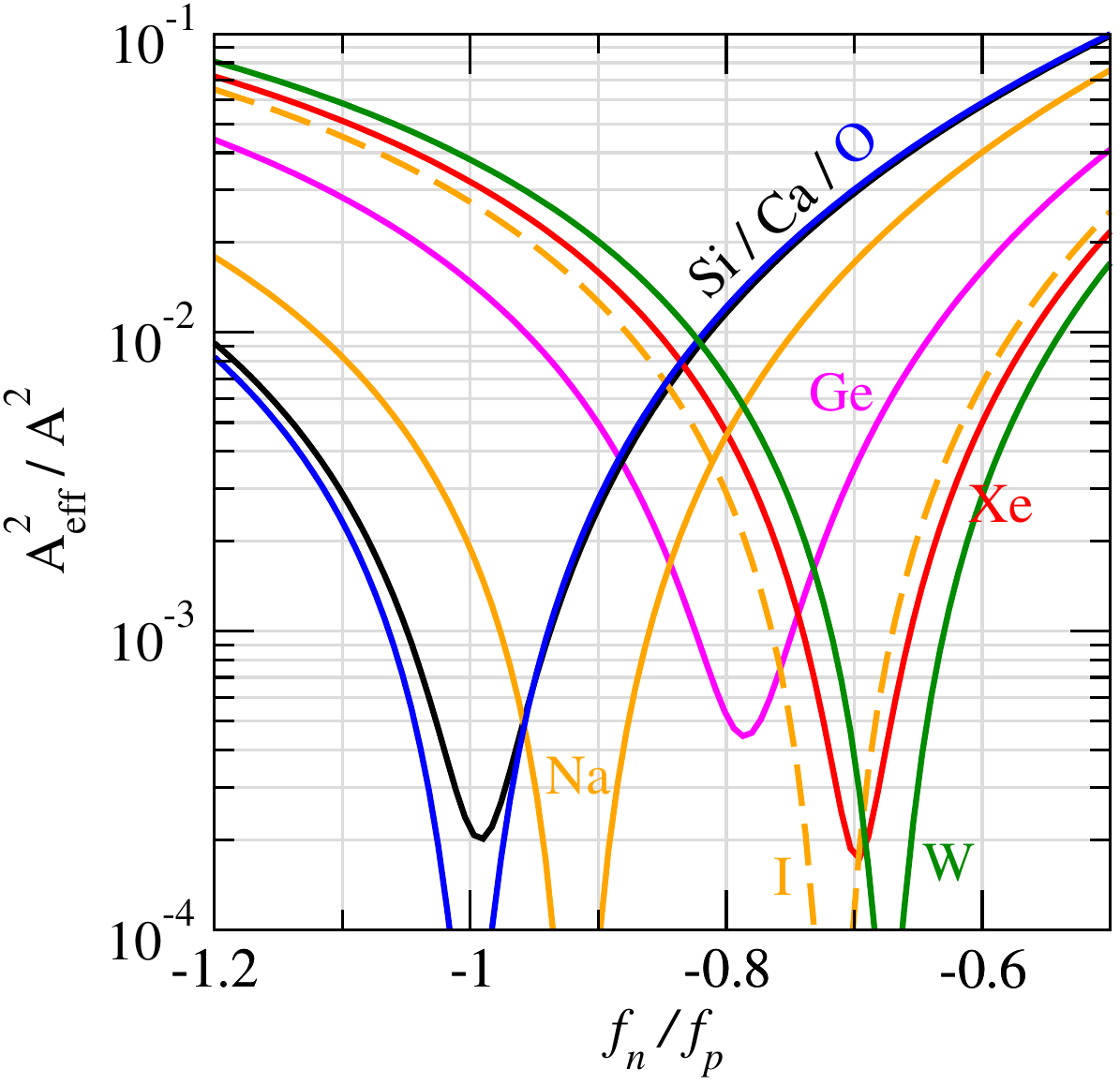} \quad
    \includegraphics[height=0.45\textwidth]{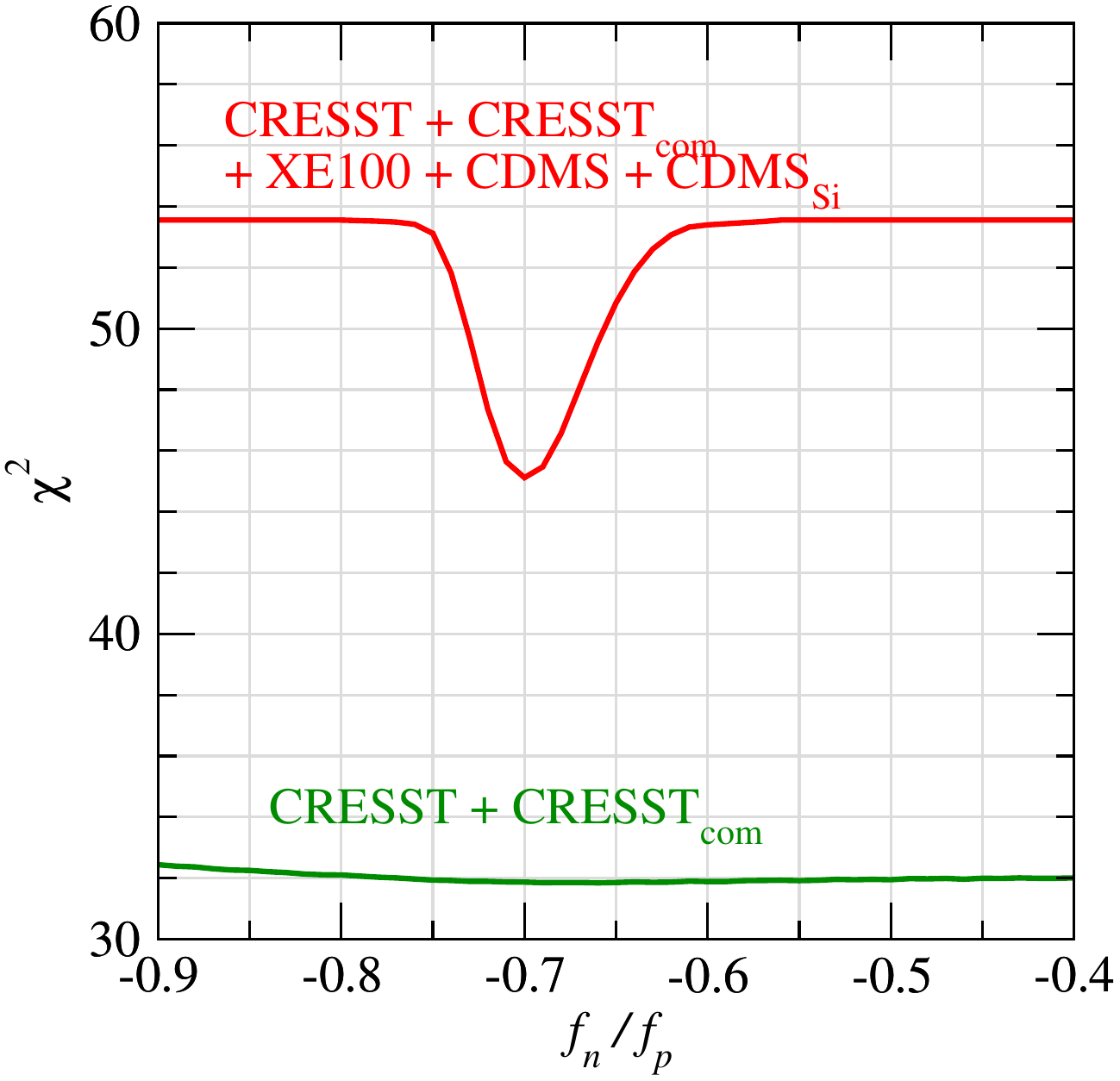}
  \mycaption{Left: the suppression factor for isospin-violating interactions
    relative to the iso-spin conserving case for various elements. $A^2_{\rm
    eff}$ is defined in eq.~\eqref{eq:Asq_eff}. Right: The total $\chi^2$ for
    CRESST (recent data from \cite{Angloher:2011uu} combined with the
    constraint from the commissioning run following \cite{Brown:2011dp}) and
    for CRESST plus data from XENON-100 \cite{Aprile:2011hi} and CDMS (Ge
    \cite{Ahmed:2009zw} + Si \cite{Akerib:2005kh} data) as a function of the
    ratio of the neutron and proton couplings.}
  \label{fig:IVDM}
\end{figure}

Finally, we return to spin-independent elastic interactions, but relax the
assumption of equal couplings of DM to protons and neutrons. That is, we will
now assume $f_p \neq f_n$ in eq.~\eqref{eq:sigmaSI}. Following
\cite{Schwetz:2011xm} we define an effective nuclear mass number $A_{\rm eff}$
such that the event rate from scattering on a given element is suppressed by
$A^2_{\rm eff} / A^2$ relative to the case $f_p = f_n$:
\begin{align}
  A^2_{\rm eff} \equiv \sum_{i \in \rm isotopes} 2 r_i [Z\cos\theta + (A_i-Z)\sin\theta]^2 \,.
  \label{eq:Asq_eff}
\end{align}
Here $\tan\theta \equiv f_n / f_p$ and $r_i$ is the relative abundances of the
$i$-th isotope in the target. Fig.~\ref{fig:IVDM} (left) shows the suppression
factor for various relevant elements.  This plot shows that for $f_n/f_p =
-0.7$ it is possible to suppress the rate on Xe and to a lesser extent also
that on Ge relative to O and Ca. Indeed, we can see from fig.~\ref{fig:IVDM}
(right) that the combined $\chi^2$ from CRESST, XENON, and CDMS has a clear
minimum at that value of $f_n/f_p$. From the left panel we expect that the
scattering rate in CRESST will be dominated by Ca and O, since W has a neutron
to proton ratio similar to Xe, leading to a suppression of the W recoil rate at
similar values of $f_n/f_p$. On the other hand, Si has a proton to neutron
ratio very similar to Ca, which implies that the bound from CDMS Si remains
important.

\begin{figure}
  \includegraphics[height=0.5\textwidth]{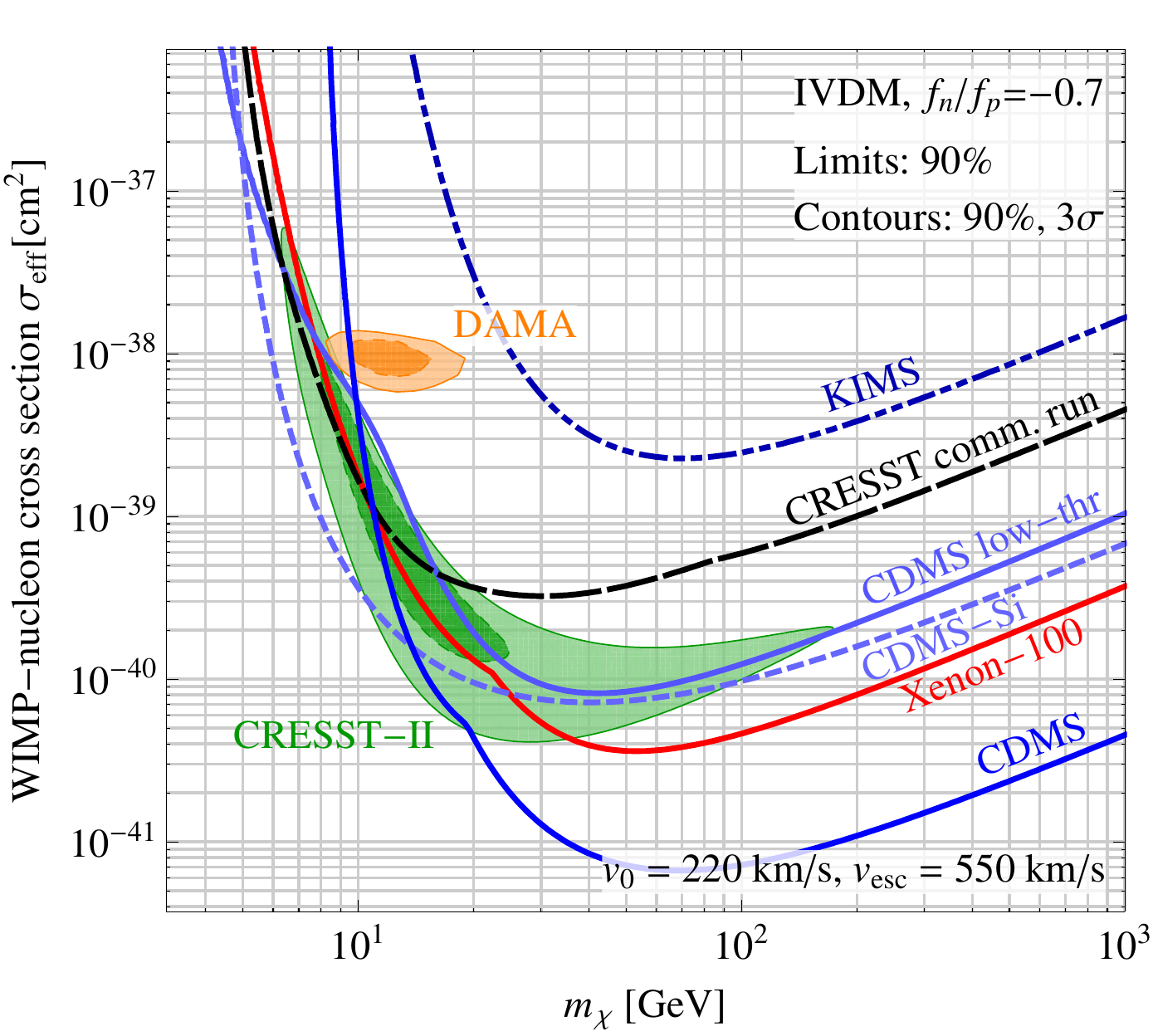} 
  \mycaption{CRESST preferred region in dark matter mass $m_\chi$ and
    average scattering cross section $\sigma_{\rm eff} = (\sigma_p+\sigma_n)/2$
    for elastically scattering isospin-violating dark matter with $f_n/f_p =
    -0.7$, and constraints from other experiments.}
  \label{fig:IVDM-reg}
\end{figure}

Fig.~\ref{fig:IVDM-reg} shows the CRESST allowed region compared to the various
limits assuming $f_n/f_p = -0.7$ (this value is chosen in order to minimize the
predicted rate in Xenon-100, thus relaxing the otherwise strongest constraint on
dark matter scattering). We observe that part of the CRESST $3\sigma$ region
remains consistent with XENON and CDMS limits at 90\%~CL.  The combined best
fit point is obtained for $m_\chi = 14.2$~GeV and $\sigma_{\rm eff} =
(\sigma_p+\sigma_n)/2 = 2.6\times 10^{-40} \, cm^2$. As expected the event rate in
CRESST is dominated by the light elements (39\% scattering on O, 61\% on Ca) with a
negligible contribution from W. The corresponding spectrum is shown in
fig.~\ref{fig:spectrum} as the red histogram. However, although there is clear
improvement with respect to the isospin-conserving case, the consistency
according to the PG test is still rather low, at 0.4\%, see table~\ref{tab:PG}.
When performing a combined analysis of CRESST+CDMS+XENON we find closed regions
only at $2\sigma$, whereas at $3\sigma$ no closed region appears, indicating
also considerable tension in the data.

Let us mention also that we do not expect further improvement by combining
isospin-violating interactions with inelastic scattering. For IVDM with $f_n /
f_p = -0.7$ scattering in the preferred parameter region occurs dominantely on
O and Ca, since W recoils are suppressed because of the proton-to-neutron ratio
$\sim 0.7$ in W. Inelastic scattering, on the other hand, occurs exclusively
on W, since scattering on the lighter targets is kinematically forbidden.
Therefore, iDM and IVDM are not compatible with each other as an explanation
for the CRESST signal.

\section{Conclusions}
\label{sec:conclusions}

In this paper, we have studied the compatibility of the results from various
direct dark matter detection experiments, including in particular new results
from CRESST, which show a $4\sigma$ excess of events above known backgrounds.
We have found that none of the considered dark matter models---elastic and
inelastic spin-independent scattering as well as isospin-violating dark
matter---can simulatenously explain the positive signals from CRESST, DAMA and
CoGeNT, while satisfying constraints from other experiments. The CRESST signal
alone, however, is consistent with exclusions from other experiments in the
case of inelastically scattering DM and DM with isospin-violating couplings to
visible matter, though mild tension remains.  The parameter
region preferred by current CRESST data and the null result from an earlier
commissioning run of the experiment are consistent at the 10\% level, as
estimated using the parameter goodness of fit method (for standard assumptions
on the DM velocity profile).

We have also briefly commented on a recent reassessment of the CoGeNT data, which
indicates that a yet unknown fraction of the excess events in CoGeNT is due to
surface backgrounds, and on a new measurement of the Na quenching factors in DAMA,
which indicates that they might be much smaller than previously thought. We have
demonstrated that a lower signal rate in CoGeNT could make their results more
consistent with the null results from XENON-100 and CDMS. The modified quenching
factors in DAMA, if confirmed, would on the other hand increase the tension
between DAMA and the null results.

We conclude that the landscape of dark matter direct detection is evolving
rapidly, and there is currently no known way of explaining all the positive
signals (DAMA, CoGeNT and CRESST) simultaneously with all the null results.
It is therefore crucial that these signals, but also the null results, are
scrutinized carefully, and experimental cross-checks are carried out to test
them as model-independently as possible, for instance by using the same target
material in several experiments.

\bigskip

 {\it Note added:} During completion of this work we learned that
the inelastic DM solution for CRESST has been independently obtained by
N.~Weiner~\cite{Weiner:2011talk}.

\acknowledgments

We thank Patrick Huff, Josef Jochum, and Jens Schmaler for helpful
discussions on the CRESST data, and Sunkee Kim and Seung Cheon Kim for
useful communication regarding the KIMS analysis. The work of T.S.\ was
partly supported by the Transregio Sonderforschungsbereich TR27 ``Neutrinos
and Beyond'' der Deutschen Forschungsgemeinschaft. Fermilab is operated by
Fermi Research Alliance, LLC, under Contract DE-AC02-07CH11359 with the
United States Department of Energy.

\bibliographystyle{my-h-physrev.bst}
\bibliography{./refs}

\end{document}